\documentclass[longbibliography, nofootinbib]{revtex4-2}
\usepackage{bm}
\usepackage{amsfonts}
\usepackage{latexsym}
\usepackage{graphicx}
\usepackage{amsmath}
\usepackage{palatino}
\usepackage{comment}
\usepackage{textcomp}
\usepackage{float}
\usepackage{booktabs}
\usepackage{dcolumn}
\usepackage{ragged2e}
\usepackage[hyperfootnotes=false]{hyperref}
\hypersetup{colorlinks=true}
\usepackage{xcolor}
\usepackage{orcidlink}
\usepackage{epsfig}
\usepackage{tcolorbox}
\usepackage{caption}
\usepackage{subcaption}
\usepackage{commath}
\usepackage{enumitem}
\def \nn  {\nonumber}

\allowdisplaybreaks[1]

\begin{document}
\title{Thermodynamics of sign-switching dark energy models}

\author{David Tamayo$^{1,2}$ \orcidlink{0000-0001-9079-8893}}
\email{david.tr@piedrasnegras.tecnm.mx}
\affiliation{$^1$Instituto Tecnológico de Piedras Negras, 26080, Piedras Negras, Mexico}
\affiliation{$^2$Instituto de Astrof\'{\i}sica e Ci\^encias do Espa{\c c}o, Universidade do Porto, CAUP, 4150-762, Porto, Portugal}

\begin{abstract}

We perform a comprehensive thermodynamic analysis of three sign-switching dark energy models in a flat FLRW cosmology: graduated dark energy (gDE), sign-switching cosmological constant ($\Lambda_s$), and smoothed sign-switching cosmological constant ($\Lambda_t$). 
We systematically derive key cosmological thermodynamic quantities—horizon temperature, horizon entropy, internal entropy, total entropy, and their first and second derivatives—using the Generalised Second Law (GSL) as the fundamental evaluation criterion. 
We first confirm the compliance of the $\Lambda$CDM model with the GSL, establishing a baseline for comparison. 
We find that despite their unconventional negative-to-positive energy density transitions, both $\Lambda_s$ and $\Lambda_t$ remain thermodynamically consistent. 
In contrast, gDE exhibits significant issues: divergences in its equation-of-state lead to infinite horizon temperature and entropy derivatives; and asymptotically, the horizon temperature diverges while entropy approaches zero, causing entropy reduction and violating the GSL. 
We highlight a key insight: models with divergences in the product of the dark energy equation-of-state parameter and its energy density ($w_x\Omega_x$) inevitably produce thermodynamic inconsistencies in standard cosmology. 
This thermodynamic approach provides a complementary criterion alongside observational constraints for evaluating the physical viability of cosmological models.\\

\end{abstract}

\maketitle
\tableofcontents

\section{Introduction}

The Lambda Cold Dark Matter ($\Lambda$CDM) model, complemented by the inflationary paradigm, remains the most successful cosmological model, accurately describing the dynamics and large-scale structure of the universe.
Nevertheless, in addition to its long-standing theoretical issues like the cosmological constant problem or the nature of dark energy and dark matter \cite{Coley:2019yov}, in the past decade persistent tensions of varying significance have emerged between some existing data sets \cite{Perivolaropoulos:2021jda}. 
Probably the most highlighted tension reported so far is the significant deficiency in the Hubble constant $H_{0}$ value predicted by the cosmic microwave background (CMB) Planck data \cite{Planck:2015fie, Planck:2018vyg} using the base $\Lambda$CDM model when compared with the values by direct model-independent local measurements \cite{Riess:2016jrr, Riess:2018byc, Riess:2019cxk, Freedman:2019jwv}. 
A plethora of attempts to alleviate the $H_{0}$ tension has appeared in recent years, such as: early dark energy \cite{Poulin:2018cxd}, modified gravity \cite{Escamilla:2024xmz}, decaying dark matter \cite{Vattis:2019efj}, phantom dark energy \cite{DiValentino:2020naf}, new physics in neutrinos \cite{Kreisch:2019yzn}, late-time transition in dark energy \cite{Alestas:2021xes}, interacting dark energy \cite{Pan:2023mie}, running vacuum \cite{SolaPeracaula:2023swx}, Tsallis entropy \cite{Basilakos:2023kvk}.
For reviews see \cite{DiValentino:2021izs, Hu:2023jqc, DiValentino:2024yew}.

One striking proposal, outside the mainstream but effectively alleviating the $H_0$ tension, suggests that dark energy density need not remain strictly positive and can become negative or vanish at finite redshift \cite{Mortsell:2018mfj, Poulin:2018zxs, Capozziello:2018jya, Wang:2018fng, Dutta:2018vmq, Ye:2020btb, Calderon:2020hoc, Sen:2021wld, DiGennaro:2022ykp, Ong:2022wrs, Malekjani:2023ple, Menci:2024rbq}, although some studies find no evidence supporting a negative cosmological constant \cite{Visinelli:2019qqu}.
The possibility of a negative energy density of the dark energy started to gain relevance after the measurement of the Ly-$\alpha$ forest of BAO by the BOSS collaboration \cite{BOSS:2014hwf}.
They reported a clear preference for $\Lambda>0 $ for $z<1$, but a preference for $\Lambda<0$ for $z>1.6$ and argued that the Ly-$\alpha $ data of $z\approx 2.34$ can be fit to a non-monotonic evolution of $H(z)$, which is difficult to achieve in the framework of general relativity in any dark energy model with strictly non-negative energy density \cite{BOSS:2014hhw}.
In line with this, it was argued that the Ly-$\alpha$ data can be accommodated by a modified gravity model that alters $H(z)$ itself, and also that a further tension relevant to the Ly-$\alpha $ data can be alleviated in models in which $\Lambda$ is dynamically screened, implying an effective dark energy passing below zero and concurrently exhibiting a pole in its equation-of-state (EoS), at $z\sim 2.4$. 
Based on the observational evidence mentioned above, in recent years, a class of dark energy models has been proposed, dubbed as sign-switching cosmological constant ($\Lambda_s$), that relax the cosmological tensions and whose main characteristic is to take into account the possibility that the Universe went through an abrupt anti-de Sitter vacua to de Sitter vacua transition characterized by a sign-switching cosmological constant (negative to positive) $\Lambda<0\; \rightarrow \; \Lambda>0$ at $z \sim 2$, and after testing with observational data they give promising results \cite{Acquaviva:2021jov, Akarsu:2021fol, Akarsu:2022typ, Akarsu:2023mfb, Paraskevas:2024ytz, Akarsu:2024eoo}.
Related to the $\Lambda_s$ model, the graduated dark energy (gDE) model proposes a rapid and smooth transition from negative to positive values of the dark energy density that can simultaneously address the $H_0$ and BAO-Ly-$\alpha$ discrepancies \cite{Akarsu:2019hmw}.
A complementary line of work shows that sign-switching dark energy can arise in both phenomenological and theoretical settings. In interacting or dynamical vacuum models, the wXCDM framework features a transition between a phantom-like component with negative energy density and a quintessence-like component with positive energy density \cite{Grande:2006nn, Gomez-Valent:2024tdb, Gomez-Valent:2024ejh}. From a more fundamental perspective, the existence of two distinct dark energy phases has also been demonstrated in the stringy Running Vacuum model, where transient domains of phantom matter may form as the Universe approaches a de Sitter phase \cite{Mavromatos:2020kzj, Mavromatos:2021urx}. These results support scenarios in which the dark energy density changes sign.

Typically, a dynamical dark energy is modelled as a perfect fluid with pressure $p$ and energy density $\rho$ related by an EoS $p = w(a) \rho$, where $w(a)$ captures the particular features of a specific dark energy model.
The parameters of the dark energy models are then constrained by observations of different nature, usually CMB, BAO and SNIa.
Despite the versatility of dynamical dark energy models to describe the cosmic expansion, alleviate cosmological tensions and pass the observational filter, it must be verified that these models do not conflict with fundamental physics, such as thermodynamics.
Beyond observational constraints, it is worthwhile to extend the analysis of the dark energy into the realm of thermodynamics \cite{Silva:2013ixa, Barboza:2015bha, Cardone:2016ewm, Duarte:2018gmt, SolaPeracaula:2019kfm, Ramirez:2020ldb}.
The interesting connection between gravity and thermodynamics is well illustrated in the case of black holes, where in their semiclassical description of their physics, black holes behave as black bodies emitting thermal radiation and have both a temperature \cite{Hawking:1975vcx} and an entropy \cite{Bekenstein:1973ur}.
According to the Hawking-Bekenstein formula, the entropy of the event horizon of a black hole is proportional to its area (along with the Planck and Boltzmann constants).
Furthermore, it has also been shown that black holes in contact with their own radiation satisfy the second law of thermodynamics \cite{Bekenstein:1974ax}.

These results have been generalised in the cosmological context.
If the universe is considered as an isolated system and filled with a cosmic fluid bounded by a horizon, then in analogy to the black hole thermodynamics it is possible to define an entropy of the whole system.  
In cosmology, the horizon usually considered is the apparent horizon, the dynamic boundary beyond which light or signals from distant regions cannot reach an observer due to the accelerating expansion of the universe, because it always exists in FLRW universes, something that is generally not true for the particle horizon and the future event horizon \cite{Wang:2005pk, Cai:2008gw}. 
The entropy of the universe is described by the Generalised Second Law (GSL) of thermodynamics which states that the total entropy of the universe $S_{\text{tot}}$ is the sum of the entropy of the horizon  $S_h$ itself with the entropy of the cosmic components  $S_{\text{in}}$ within it.
The GSL was first stated for black holes by Bekenstein pioneer work \cite{Bekenstein:1974ax}, and then Davies \cite{Davies:1987ti, Davies:1988dk} explored the event horizon within the context of FLRW space, particularly in the context of General Relativity with a perfect fluid acting as the source.
If considered as an isolated thermodynamical system, it should be expected that the universe evolve naturally towards a thermodynamic equilibrium, and therefore, the total entropy function $S(x)$ (where $x$ is a relevant variable describing the temporal evolution of the system), should be a positively definite quantity that increases with convex concavity \cite{callen1960, Padmanabhan:2003gd, Wang:2005pk, Faraoni:2015ula}.
In other words, any physically viable cosmological model must satisfy the two conditions $dS(x)/dx\geq 0$ and, in the far future, $d^2S(x)/dx^2 < 0$.
The GSL gives a possible method to discriminate between observationally adequate dark energy models, but not necessarily thermodynamically plausible by calculating the total entropy of the Universe (including both the horizon contribution and the matter-energy contained within it) and verify that its first and second derivatives behave correctly
\cite{Izquierdo:2005ku, Gong:2006ma, Radicella:2011qpl, Cardone:2016ewm, daSilva:2020ylz, Komatsu:2023wml, Cruz:2023xjp, Odintsov:2024hzu}. 

A dark energy model is generally considered well-motivated if it fits reasonably well with observational data and also alleviates some of the problems of cosmology, such as the cosmological constant problem or the Hubble tension.
This is fine, but perhaps not sufficient.
It is worthwhile to approach the dark energy problem from viewpoints beyond phenomenological models or observational reconstructions.
Since we do not yet have an underlying theory for the nature of dark energy, thermodynamics provides a useful approach in this regard, as it studies systems in a general way without considering their microphysical constitution.
Regardless of the specific microphysical nature dark energy may have, it must obey the laws of thermodynamics.
Therefore, bounds can be placed on dark energy models by analysing their thermodynamic properties.

This article is structured as follows. 
In Section~\ref{sec:models}, we present the three sign-switching dark energy models analysed: graduated dark energy, abrupt sign-switching cosmological constant, and its smoothed variant. 
Section~\ref{sec: Cosmological thermodynamics} establishes the cosmological thermodynamic formalism and the criteria defined by the Generalised Second Law. 
In Section~\ref{sec:LCDM_thermo}, we apply this framework to the standard $\Lambda$CDM model, which serves as our baseline for comparison. 
The thermodynamic consistency of the sign-switching models is then thoroughly examined in Section~\ref{sec:sign_switching_thermo}. 
Finally, Section~\ref{sec:conclusions} summarises our findings and outlines directions for future research.

\section{Sign-switching dark energy models}\label{sec:models}

Before introducing the dark energy models under study, we first establish the theoretical framework.  
Within the framework of General Relativity, we consider a homogeneous, isotropic, and spatially flat universe described by the FLRW metric.  
The evolution of the cosmic background is governed by the Friedmann equations:
\begin{eqnarray}
H^2 &=& \frac{8 \pi G}{3} \sum_i{\rho_i}, \label{Friedmann 1}\\
\dot{H} +H^2 &=& -\frac{4 \pi G}{3} \sum_i\left[{\rho_i} +3{p_i}\right],
\label{Friedmann 2}
\end{eqnarray}
where $H = \dot{a}/a$ is the Hubble expansion rate, $a$ the scale factor (with present day value $a_0 = 1$), the dot denotes derivative with respect to cosmic time, and the sum is over the cosmic components.
Hereafter we will use $(r, m, x)$ to refer to the radiation, matter (pressureless baryonic and dark matter), and dark energy terms.
 
The continuity equation, assuming no interaction between fluids, is
\begin{eqnarray}
    \dot{\rho}_i +3H(1+w_i)\rho_i =0,
    \label{continuity}
\end{eqnarray}
where $w_i$ is the EoS parameter of the $i$-th term. 
Setting $w_r = 1/3$ for radiation and $w_m = 0$ for matter, then from the first Friedmann equation we have
\begin{equation}
E = \frac{H}{H_0} = \sqrt{\Omega_{m0} a^{-3} + \Omega_{r0} a^{-4} +\Omega_{x0} F_x(a)}
\label{E}
\end{equation}
where
\begin{equation}
\Omega_{i0} = \left.\frac{\rho_i}{\rho_{crit}}\right\rvert_{a=1} = \frac{8 \pi G}{3 H_0^2}  \rho_{i0},
\end{equation}
is the density parameter of the $i$-th component at the present time, while $F_x(a) = \rho_x (a)/\rho_{x0}$ is a general function characterising each specific dark energy model, which must be determined by integrating equation \eqref{continuity} for a given dark energy EoS.

\subsection{Graduated dark energy}

In this subsection we will summarize the theoretical background of the \textit{graduated dark energy} (gDE), proposed by Akarsu et al. in \cite{Akarsu:2019hmw}.
The gDE is a dark energy model that can be viewed as a dynamical deviation from the null inertial mass density $\rho_{\textbf{inert}}$ described by the conventional vacuum (cosmological constant), $\Lambda$ in the form $\rho_{\textbf{inert}} \propto \rho^{\lambda}$, where $\rho_{\textbf{inert}} = \rho_x +p_x$ is the inertial mass density (energy density plus pressure) of the vacuum (for $\Lambda$CDM $\rho_{\textbf{inert}}= 0$) and $\lambda$ a constant. 
This allows the energy density to spontaneously switch sign at a characteristic redshift ($z \sim 2.3$). 
This behaviour provides a natural explanation for late-time cosmic acceleration and addresses key tensions in the $\Lambda$CDM model.
Observational analysis show that gDE significantly improves the fit to data compared to $\Lambda$CDM, achieving better agreement with independent $H_0$ measurements and resolving discrepancies with high-precision Ly-$\alpha$ forest data. 
Furthermore, gDE suggests a transition from anti-de Sitter (AdS) to de Sitter (dS) vacua, aligning with string theory frameworks, which could have profound implications for theoretical physics and the nature of dark energy.

The proposed gDE is a dark energy model with an inertial mass density exhibiting power-law dependence to its energy density as follows,
\begin{eqnarray}
    \rho_{\textbf{inert}} = \gamma \rho_0\left(\frac{\rho_x}{\rho_{x0}}\right)^\lambda,
\end{eqnarray}
where $\rho_{x0}$ is the present-day gDE energy density value, and the parameters $\gamma$ and $\lambda$ are real constants. 
The gDE EoS parameter $w_x = p_x/\rho_x = -1 + p_x/\rho_{\textbf{inert}}$ is
\begin{eqnarray}\label{w rho1}
    w_x = -1 +\gamma\left(\frac{\rho_x}{\rho_{x0}}\right)^{\lambda-1},
\end{eqnarray}
where $\gamma = 0$ corresponds to $\Lambda$CDM ($w_x = -1$).
Substituting equation \eqref{w rho1} into the continuity equation \eqref{continuity}, we can solve to obtain the explicit form of the gDE energy density and the EoS parameter:
\begin{eqnarray}
    \rho_{x} &=& \rho_{x0} [1 +3\gamma(\lambda-1)\ln a]^{\frac{1}{1-\lambda}} \label{rhox1}, \\
    w_{x} &=& -1 +\frac{\gamma}{1 +3\gamma(\lambda-1)\ln a}. \label{w rho}
\end{eqnarray}
When the parameters $\gamma$ and $\lambda$ are chosen appropriately, gDE can describe a transition from $\rho_x<0$ to $\rho_x>0$, establishing itself as a phenomenological dark energy model in which a smooth function represents the cosmological constant switching signs at a specific redshift.
We consider the case where the energy density starts below zero at high redshifts and, after the sign switch, becomes positive and approaches the cosmological constant.
To analyse this, note that the energy density in \eqref{rhox1} follows a power-law behaviour, defined as $\rho_x/\rho_{x0} = x^y$, where $\rho_{x0} > 0$, $x = 1 + 3\gamma(\lambda-1) \ln a$, and $y = \frac{1}{\lambda-1}$.
Unless $\gamma=0$ (cosmological constant) or $\lambda=1$ (constant EoS parameter dark energy), $x$ changes sign at:
\begin{equation}
    a_* = \exp \left(-\frac{1}{3}\frac{1}{\gamma(\lambda-1)}\right).
\end{equation}
It is important to note that the EoS parameter \eqref{w rho} diverges at $a=a_*$, corresponding precisely to the point at which the energy density changes sign.
If $\gamma(\lambda-1) > 0$, the sign change occurs in the past ($a_* < 1$), while if $\gamma(\lambda-1) < 0$, it occurs in the future. 
Furthermore, if $\lambda > 1$ (then $y < 0$), $\rho \rightarrow \pm \infty$ as $a \rightarrow a_*$, meaning the energy density diverges at the sign switch; on the other hand, if $\lambda < 1$ (then $y > 0$), $\rho \rightarrow 0$ as $a \rightarrow a_*$.
Thus, it is necessary to fix the parameters as follows:
\begin{equation}
     \lambda < 1, \quad \gamma<0, 
\end{equation}
to restrict the model to the desired scenario. 
To avoid possible sign ambiguities, it is convenient to use the sign function $\rho_x/\rho_{x0} = x^y = \text{sgn}(x)|x|^y$, allowing us to rewrite the gDE energy density \eqref{rhox1} as:
\begin{eqnarray}
    \rho_x = \rho_{x0}\, \text{sgn}(1-\Psi \ln a)|1-\Psi \ln a|^{\frac{1}{1-\lambda}}, \label{rhox}
\end{eqnarray}
with $\Psi = -3\gamma(\lambda -1) < 0$. 

Having established the dark energy model described by \eqref{rhox}, the dimensionless Hubble parameter $E$ can be written as:
\begin{eqnarray}
    E = \sqrt{\Omega_{r0} a^{-4} +\Omega_{m0} a^{-3} +\Omega_{x0}\, \text{sgn}(1-\Psi \ln a)|1-\Psi \ln a|^{\frac{1}{1-\lambda}}}
\end{eqnarray}

\subsection{Sign-switching $\Lambda$}

The \textit{sign-switching $\Lambda$} model, or $\Lambda_s$, inspired by gDE, was introduced in \cite{Akarsu:2021fol}. 
This model considers a cosmological constant, $\Lambda$, that transitions from negative values in the early universe to positive values in the present day, occurring at a specific critical redshift, $z_*$.
The $\Lambda_s$ model encapsulates the phenomenology of gDE within a simpler and more observationally accessible framework. 
It alleviates the $H_0$ tension and resolves discrepancies with Lyman-alpha BAO measurements, while maintaining consistency with Planck CMB data. 
Additionally, it reduces the $S_8$ tension and aligns more closely with the tip of the red giant branch (TRGB) $H_0$ measurements.
However, the model also presents challenges, such as reconciling its predictions with galaxy BAO measurements and refining the theoretical motivation behind the sign-switch mechanism. 
The $\Lambda_s$ is described as:
\begin{equation}\label{Lambda s}
\Lambda_s \equiv \Lambda_{s0} \, \text{sgn}[z_* - z],
\end{equation}
where $\Lambda_{s0} > 0$ is the magnitude of the present-day cosmological constant and $\text{sgn}(x)$ is the sign function, taking values $-1$, $0$, and $1$ for $x < 0$, $x = 0$, and $x > 0$, respectively.

The energy density and EoS parameter for the sign-switching $\Lambda_s$ dark energy model are given by:
\begin{eqnarray}\label{rhox ss}
\rho_x  &=& \rho_{s0} \, \text{sgn}\left[\frac{1}{a_*} - \frac{1}{a}\right],\\
w_x &=& -1
\end{eqnarray}
where the sign-switch transition scale factor is $a_* = \frac{1}{1 + z_*}$.
Substituting equation \eqref{rhox ss} into the Friedmann equation \eqref{E} we obtain:
\begin{eqnarray}
    E = \sqrt{\Omega_{r0} a^{-4} +\Omega_{m0} a^{-3} +\Omega_{x0}\, \text{sgn}\left[\frac{1}{a_*} - \frac{1}{a}\right]}
\end{eqnarray}

\subsection{Sign-switch transition with $\tanh$}

The transition of the cosmological constant, $\Lambda_s$, from negative to positive values is described as abrupt, using the sign function (see equation \eqref{Lambda s}). 
However, this should be understood as an idealized representation of a rapid transition, not strictly instantaneous. 
To provide a more realistic description, the transition can be smoothed using a hyperbolic tangent function, $\tanh(x)$, which avoids mathematical discontinuities \cite{Akarsu:2022typ}. 
The smoothed transition is expressed as:
\begin{eqnarray}\label{Lambda tahn}
\Lambda_t &=& \Lambda_{s0} \, \tanh[\eta(z_* - z)],
\end{eqnarray}
where, the parameter $\eta > 0$ controls the rapidity with which the cosmological constant transitions from negative to positive values. 
Large values of $\eta$ correspond to an abrupt transition (similar to the $\Lambda_s$ model), while smaller values yield a smoother transition. 
The sign-switch transition redshift $z = z_*$ and $\Lambda_{s0} > 0$ are the same constants used in the sign-switching $\Lambda$ model. 
In the limit $\eta \to \infty$, this expression recovers the abrupt behaviour of the $\text{sgn}(x)$ function.
By employing a continuous function such as $\tanh(x)$, the model gains flexibility and mathematical smoothness, making it more suitable for observational and theoretical analyses. 
This refinement also enables the exploration of scenarios where the transition occurs over a finite interval rather than instantaneously.

The energy density for the smoothed transition in the $\tanh$ dark energy model is given by:
\begin{eqnarray}\label{rhox tahn}
\rho_x  = \rho_{x0} \, \tanh\left[\eta\left(\frac{1}{a_*} - \frac{1}{a}\right)\right].
\end{eqnarray}
To determine the EoS parameter, we rewrite the conservation equation \eqref{continuity} in the form $w_x = -1 -\frac{a}{3\rho_x}\frac{d\rho_x}{da}$.
Integrating this expression yields:
\begin{eqnarray}\label{wx tahn}
w_x  = -1 -\frac{2 \eta}{3 a \sinh{\left(2 \eta \left(\frac{1}{a_*} - \frac{1}{a}\right) \right)}}.
\end{eqnarray}
Finally, substituting \eqref{rhox tahn} into the Friedmann equation \eqref{E}, we obtain:
\begin{eqnarray}
    E = \sqrt{\Omega_{r0} a^{-4} + \Omega_{m0} a^{-3} + \Omega_{x0}\, \tanh\left[\eta\left(\frac{1}{a_*} - \frac{1}{a}\right)\right]}.
\end{eqnarray}

\renewcommand{\figurename}{Figure}
\begin{figure}[ht!]
    \centering
    \includegraphics[width=0.45\linewidth]{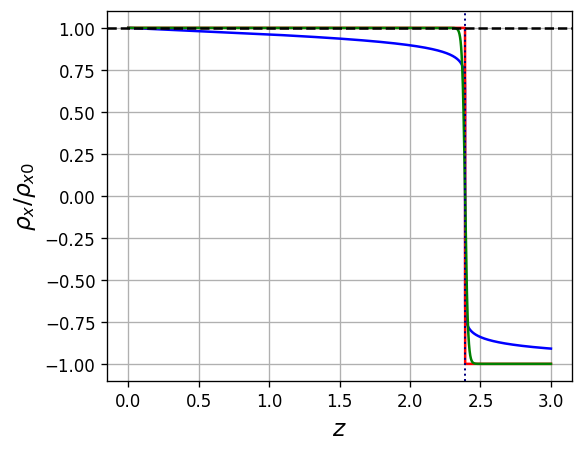}
    \includegraphics[width=0.45\linewidth]{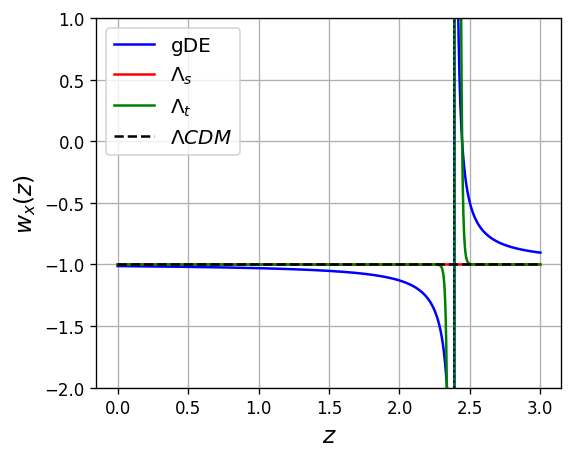}
    \includegraphics[width=0.45\linewidth]{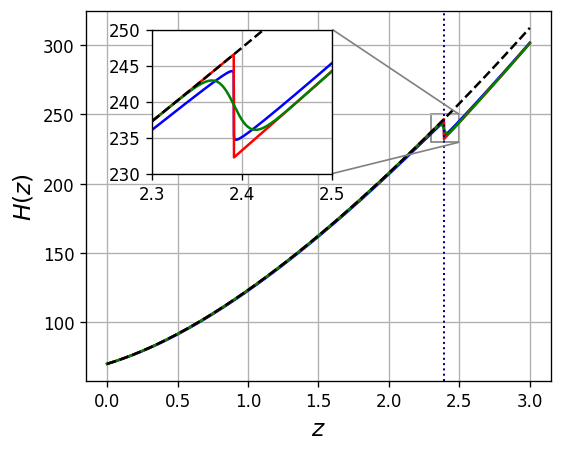}
    \caption{Normalized dark energy density, EoS parameter, and Hubble function for sign-switching dark energy models: gDE (blue), $\Lambda_s$ (red), $\Lambda_t$ (green); and $\Lambda$CDM (dashed black). For the gDE model, the parameters are $\gamma = -0.013$ and $\lambda = -20$, resulting in a sign-switch transition redshift of $z_* = 2.39$, which is also used for both $\Lambda_s$ and $\Lambda_t$. For the $\Lambda_t$ model, the smoothing parameter $\eta=50$ is arbitrary and intended solely to illustrate a rapid but not strictly instantaneous transition.}
    \label{fig:rho w H}
\end{figure}

In Fig. \ref{fig:rho w H}, the behaviour of the three sign-switching dark energy models is illustrated by plotting the normalized dark energy density $\rho_x/\rho_{x0}$, the EoS parameter $w_x$, and the Hubble function $H(z)$. 
All models share the same sign-switch transition redshift, $z_* = 2.39$, obtained from the best fit of the gDE parameters $\gamma = -0.013$ and $\lambda = -20$, as shown in \cite{Akarsu:2019hmw}. 
The cosmological background parameters used in all cases are $H_0 = 70$ km s${}^{-1}$ Mpc${}^{-1}$, $\Omega_{m0} = 0.3$, $\Omega_{r0} = 0.001$, and $\Omega_{x0} = 1 - \Omega_{m0} - \Omega_{r0}$.
For the energy density plot, we observe that the smoothest transition corresponds to gDE, with $\Lambda_t$ showing an intermediate behaviour, and $\Lambda_s$ exhibiting an abrupt (discontinuous) transition. 
A similar trend is observed in the Hubble function.
On the other hand, for the EoS parameter $w_x$, both gDE and $\Lambda_t$ display a divergence at $z_*$, with $\Lambda_t$ showing a sharper divergence. 
This singularity observed in $w_x$ for the gDE and $\Lambda_t$ is a mathematical feature intrinsic to these models, which does not necessarily represent a physical pathology, as is not a directly observable quantity.
By construction, $\Lambda_s$ maintains the standard form $w_x = -1$ throughout. 
Although the singularity in the EoS parameter may appear pathological, it is important to note that $w_x$ represents the ratio of two physical quantities and is not directly observable. 
Discontinuous and divergent EoS parameters for dark energy have been reported in other plausible models, as discussed in \cite{Sahni:2014ooa, Akarsu:2019ygx, Bag:2021cqm, Ozulker:2022slu}.

\section{Cosmological thermodynamics} \label{sec: Cosmological thermodynamics}

For the thermodynamic analysis, we will calculate the entropy and cosmological temperature of the three dark energy models presented in the previous section. 
The GSL will be applied to evaluate their thermodynamic feasibility.
The thermodynamic constraints imposed by the GSL are as follows: 
\begin{enumerate}[label=\alph*)]
    \item The total entropy is the sum of the apparent horizon entropy and the entropy of the cosmic fluids contained within the horizon.
    \item The total entropy must remain constant or increase over time.
    \item In the distant future, the entropy must continue increasing but at a decreasing rate.
\end{enumerate}

These conditions can be expressed mathematically as: 
\begin{eqnarray} 
S_{\text{tot}} &=& S_h + S_{\text{in}}, \label{S tot}\\
S_{\text{tot}}^{\prime}(a) &\geq& 0, \label{S' tot}\\
S_{\text{tot}}^{\prime \prime}(a) &\leq& 0, \quad a \gg 1. \label{S'' tot}
\end{eqnarray} 
where the prime denotes the derivative with respect to the scale factor.

\subsection{Apparent horizon entropy and temperature}\label{subsec S hor}

As a first step, the apparent horizon entropy must be calculated. 
This is achieved by replacing the black hole event horizon with the cosmological apparent horizon in the entropy formula.
Writing the FLRW line element as $ds^2 = h_{ab} dx^a dx^b + \tilde{r}^2 d\Omega^2$, where $\tilde{r} = a(t) r$ is the comoving radius and $h_{ab} = \text{diag}(-c^2,a(t)^2)$, the apparent horizon is defined by the condition $h^{ab}\partial_a\tilde{r} \partial_b\tilde{r} =0$ \cite{Wang:2005pk}. 
Therefore, the horizon radius for a flat FLRW universe turns out to be $\tilde{r}_h = c/H$, so that the area is ${\cal{A}} = 4 \pi \tilde{r}_h^2$. Since the entropy is proportional to ${\cal{A}}$, we obtain:
\begin{equation}
S_h = \frac{k_B c^3}{G \hbar} \frac{\mathcal{A}}{4} = \frac{k_B c^5 \pi}{G \hbar} \frac{1}{H^2}  = \frac{\pi}{G}\frac{1}{H^2},
\label{S_h}
\end{equation}
with $k_B$ as the Boltzmann constant.
We adopt natural units $k_B = c = \hbar = 1$. 
It is important to highlight that the entropy of the horizon $S_h$ is expressed in terms of the Hubble expansion rate $H$, and therefore depends on the abundance and EoS of the cosmic components.
For mathematical convenience, we normalise with $S_{h0} = \frac{\pi}{G}\frac{1}{H_0^2}$.
We obtain the normalized horizon entropy, which is simply the inverse of the square of the dimensionless Hubble parameter:
\begin{eqnarray}
    \frac{S_h}{S_{h0}} = \left(\frac{H_0}{H}\right)^2 = \frac{1}{E^2}.
    \label{S hor}
\end{eqnarray}

For the thermodynamic study, we are interested in the change of entropy throughout cosmic evolution. 
Therefore, to study the rate of change of the horizon entropy, it is useful to use the scale factor $a$ as a dynamic variable for calculating the derivatives. 
Before calculating the derivatives, it is convenient to rewrite the second Friedmann equation \eqref{Friedmann 2} in terms of the scale factor as\footnote{Here a key point is the assumption that is valid the EoS $p_i=w_i \rho_i$ for each cosmic component.}:
\begin{eqnarray}
\frac{H'}{H} = -\frac{3}{2a}\bigl(1+\sum_i w_i\Omega_i \bigr), \label{friedmann 2 a}
\end{eqnarray}
where the sum is over the cosmic components, and the density parameters $\Omega_i = \rho_i/\rho_{crit}$ are:
\begin{eqnarray}
    \Omega_m = \frac{\Omega_{m0} a^{-3}}{E^2}, \quad \Omega_r = \frac{\Omega_{r0} a^{-4}}{E^2}, \quad \Omega_x = \frac{\Omega_{x0} F_x(a)}{E^2},
\end{eqnarray}
Now, differentiating the horizon entropy \eqref{S hor} and using \eqref{friedmann 2 a} for simplification, we obtain:
\begin{equation}\label{dS hor}
\frac{S^{\prime}_h}{S_{h0}} = \frac{3}{a E^2} \bigl(1 +\sum_{i}{w_i \Omega_i}\bigr).
\end{equation}
The last equation, \eqref{dS hor}, describes the rate of change of the horizon entropy in terms of the dimensionless Hubble parameter (background cosmic expansion), as well as the different types and abundances of the cosmic fluids (terms inside the sum).

Continuing, the second derivative of the entropy is given by:
\begin{eqnarray}\label{ddS hor}
\frac{S^{\prime \prime}_h}{S_{h0}} &=& \frac{3}{a^2 E^2}\left[3\bigl(1 +\sum_{i}{w_i \Omega_i}\bigr)^2 -\bigl(1 +\sum_{i}{w_i \Omega_i}\bigr) +a\left(\frac{1}{3}\Omega_r^{\prime} +w_x\Omega_x^{\prime}+ w_x^{\prime}\Omega_x\right)\right]\nn\\
 &=& \frac{S^{\prime}_h}{S_{h0}}\left[\frac{2 +3\sum_{i}{w_i \Omega_i}}{a} +\frac{\frac{1}{3}\Omega_r^{\prime} +w_x\Omega_x^{\prime}+ w_x^{\prime}\Omega_x}{1 +\sum_{i}{w_i \Omega_i}}\right].
\end{eqnarray}
In the last equality, the second derivative is expressed in terms of the first derivative shown in \eqref{dS hor}, weighted by a factor that depends on the abundances of the different cosmic fluids as well as their dynamics throughout cosmic history.
The approach to thermodynamic equilibrium of the horizon, $S^{\prime \prime}_h = 0$, is then driven by the EoS of the fluid components, which determines their abundance evolution and their derivatives \cite{Cardone:2016ewm}.

In the same way that a horizon entropy can be defined, a horizon temperature can also be introduced. 
The first approach to a cosmological horizon temperature is through the Gibbons–Hawking temperature, $T_{GH} = \frac{\hbar H}{2\pi k_B} = \frac{H}{2\pi}$ \cite{Gibbons:1977mu}, which indicates that the temperature is proportional to the Hubble function $H$. 
The $T_{GH}$ was derived using field theory in the de Sitter spacetime. 
In the standard cosmological model, our universe is asymptotically de Sitter rather than purely de Sitter, so $T_{GH}$ serves as an approximation for the very late universe where the cosmological constant dominates. 
Therefore, a more general definition of horizon temperature is needed. 
The horizon temperature has also been studied in the context of black holes, showing a relationship between surface gravity and temperature on a dynamic apparent horizon \cite{Hayward:1997jp, Hayward:2008jq}. 
Along these lines, a more general temperature for the cosmological horizon has been proposed, commonly known as the Kodama–Hayward temperature, defined as:
\begin{equation}\label{T H def}
    T_h = \frac{|\kappa|}{2\pi},
\end{equation}
where $\kappa$ is the surface gravity \cite{Kodama:1979vn, Akbar:2006kj, Cai:2006rs}. 
This temperature has been explored from various perspectives by several authors \cite{Tu:2017jen, Mitra:2014koq, Faraoni:2015ula, Nojiri:2022nmu, Sanchez:2022xfh, Muhsinath:2022cij, Komatsu:2023wml}.

The FLRW metric can be written in the double-null form $ds^2 = h_{ab} dx^a dx^b +\tilde{r}^2 d\Omega^2 = -2d\xi^{+}d\xi^{-} +\tilde{r}^2 d\Omega^2$, where $\partial_{\pm}=\frac{\partial}{\partial\xi^{\pm}}=-\sqrt{2}\left(\frac{\partial}{\partial t}\pm\frac{\sqrt{1-ar^{2}}}{a}\frac{\partial}{\partial r}\right)$ are future-pointing null vectors. Then, the surface gravity is defined as:
\begin{equation}
\kappa = \frac{1}{2\sqrt{-h}}\partial_{a}\left(\sqrt{-h}h^{ab}\partial_{b} \tilde{r}\right).
\end{equation}
Using the relation between surface gravity and temperature \eqref{T H def}, and applying the last equation, we obtain the temperature of the horizon \cite{Cai:2006rs, Mitra:2014koq, Tu:2017jen}:
\begin{eqnarray}\label{Kodama–Hayward temperature}
    T_h = \frac{H}{2\pi} \left(1 +\frac{\dot{H}}{2H^2}\right).
\end{eqnarray}
The normalised horizon temperature is given by:
\begin{eqnarray}\label{T horizon}
    \frac{T_h}{T_{h0}} = \frac{H}{H_0} \left(1 +\frac{\dot{H}}{2H^2}\right) = E\left[1-\frac{3}{4}\bigl(1 +\sum_i w_i\Omega_i \bigr)\right],
\end{eqnarray}
where $T_{h0} = \frac{H_0}{2\pi}$ is the Gibbons–Hawking temperature at the present time.
This shows explicitly how the horizon temperature is linked directly to the dynamics of the universe’s expansion and its material content.

\subsection{Entropy of the fluid within the horizon} \label{subsec S in}

To calculate the entropy of the cosmic fluid inside the cosmological horizon, or internal entropy for short, we start with the Gibbs equation:
\begin{eqnarray}
    T dS &=& d\left(\frac{\rho}{n}\right) + p \, d\left(\frac{1}{n}\right), \\
    \Rightarrow \quad T_{\text{in}} dS_{\text{in}} &=& dU + p \, dV, \label{1 law}
\end{eqnarray}
where $S_{\text{in}}$, $T_{\text{in}}$, $U$, $V$, and $n$ represent the entropy, temperature, internal energy, volume, and particle number density of the fluid within the horizon, respectively. 
The volume is bounded by the apparent horizon:
\begin{equation}
    V = \frac{4\pi}{3} \tilde{r}_h^3 = \frac{4\pi}{3} \frac{1}{H^3}. \label{volume}
\end{equation}
From \eqref{1 law}, we can calculate the rate of change of the internal entropy with respect to cosmic time:
\begin{equation}
    \dot{S}_{\text{in}} = \frac{1}{T_{\text{in}}} \sum_i \left[ (\rho_i + p_i) \dot{V} + \dot{\rho}_i V \right]. \label{dotSmat}
\end{equation}

To solve this equation, we need to determine the temperature of the cosmic fluids $T_{\text{in}}$ within the horizon. 
The standard approach assumes that, when nonrelativistic matter and dark energy dominate the cosmic evolution, the fluid within the horizon is approximately in thermodynamic equilibrium with the horizon. 
Consequently, the temperature $T_{\text{in}}$ is commonly approximated by the apparent horizon temperature $T_h$, i.e., $T_{\text{in}} \approx T_h$ \cite{Izquierdo:2005ku, Wang:2005pk, Gong:2006ma, Duary:2023nnf}.  
Nevertheless, this assumption is an approximation and must be treated with caution, as discussed in the literature (see for example \cite{Mimoso:2016jwg, Odintsov:2024ipb}).

We use the Friedmann equations \eqref{Friedmann 1} and \eqref{Friedmann 2} to eliminate the $(\rho_i + p_i)$ term. 
Then, substituting $T_{\text{in}} = T_h$ from \eqref{Kodama–Hayward temperature}, the volume $V$ from \eqref{volume}, its derivative, and the continuity equation \eqref{continuity} into \eqref{dotSmat}, we perform algebraic manipulations to obtain:
\begin{equation}
    \dot{S}_{\text{in}} = \frac{2\pi}{G} \frac{\dot{H}}{H^3} \left( 1 + \frac{\dot{H}}{\dot{H} + 2 H^2} \right). 
\end{equation}

This expression represents the general form of $\dot{S}_{\text{in}}$ for a flat FLRW universe. 
For convenience, we rewrite $\dot{S}_{\text{in}}$ as a function of the scale factor and divide it by $S_{h0} = \frac{\pi}{G}\frac{1}{H_0^2}$ to make it dimensionless. 
After simplification, we obtain:
\begin{eqnarray}
    \frac{S_{\text{in}}^\prime}{S_{h0}} = 2 \left(\frac{H_0^2}{H^2}\right) \frac{H^\prime}{H} \left(1 + \frac{a \frac{H^\prime}{H}}{a \frac{H^\prime}{H} + 2} \right).
\end{eqnarray}
Next, substituting $E$ from \eqref{E} and $\frac{H^\prime}{H}$ from \eqref{friedmann 2 a}, we have:
\begin{equation}
    \frac{S_{\text{in}}^\prime}{S_{h0}} = -\left[\frac{3(1 + \sum_i w_i \Omega_i)}{a E^2} \right] \left[1 - \frac{3\left(1 + \sum_i w_i \Omega_i \right)}{1 - 3\sum_i w_i \Omega_i} \right]. 
\end{equation}
Notice that the first term in brackets corresponds to $\frac{S_h^\prime}{S_{h0}}$, as shown in \eqref{dS hor}. 
Hence, can be expressed as:
\begin{equation}\label{dS in}
    \frac{S_{\text{in}}^\prime}{S_{h0}} = -\left(\frac{S_h^\prime}{S_{h0}}\right) \left[1 - \frac{3\left(1 + \sum_i w_i \Omega_i \right)}{1 - 3\sum_i w_i \Omega_i} \right].
\end{equation}

Equation \eqref{dS in} describes the rate of change of the internal entropy in a flat FLRW universe. 
It is expressed in terms of the rate of change of the horizon entropy, $S'_h$, multiplied by a factor that depends on the material content of the universe. 
The summation inside the brackets is general and applies to any combination or type of cosmic fluids.
This dependence arises from the assumption that the cosmic fluids share the same temperature as the horizon, i.e., $T_{\text{in}} = T_h$. 
A significant implication of this relation is that a constant horizon entropy, $S'_h = 0$, necessarily implies constant internal entropy, $S'_{\text{in}} = 0$.

Differentiating equation \eqref{dS in} with respect to the scale factor, we obtain the second derivative of the internal entropy.
This second derivative can be expressed in terms of the second and first derivatives of the horizon entropy, weighted by the abundances and dynamics of the cosmic fluids:
\begin{equation}\label{ddS in}
    \frac{S_{\text{in}}^{\prime\prime}(a)}{S_{h0}} = -\left(\frac{S_h^{\prime\prime}}{S_{h0}}\right) \left[1 - \frac{3\left(1 + \sum_i w_i \Omega_i \right)}{1 - 3\sum_i w_i \Omega_i} \right] 
    + 12 \left(\frac{S_h^\prime}{S_{h0}}\right) \left[\frac{\frac{1}{3}\Omega_r^\prime + w_x \Omega_x^\prime + w_x^\prime \Omega_x}{(1 - 3\sum_i w_i \Omega_i)^2}\right]. 
\end{equation}

\subsection{Total cosmological entropy}

The GSL states that the entropy of a cosmological system is the sum of the horizon entropy and the internal entropy \eqref{S tot}, $S_{\text{tot}} = S_h + S_{\text{in}}$.  
In the previous sections, \ref{subsec S hor} and \ref{subsec S in}, we calculated the first and second derivatives with respect to the scale factor of the horizon entropy $S_h$ and the internal entropy $S_{\text{in}}$.  
Thus, the first derivative $S_{\text{tot}}^\prime = S^\prime_h + S^\prime_{\text{in}} $ is:
\begin{eqnarray}\label{dS tot}
    \frac{S_{\text{tot}}^\prime}{S_{h0}} = \frac{9}{a E^2}\frac{(1+\sum_i w_i\Omega_i)^2}{1-3\sum_i w_i\Omega_i}. 
\end{eqnarray}

And the second derivative:
\begin{eqnarray}\label{ddS tot}
    \frac{S_{\text{tot}}^{\prime\prime}}{S_{h0}} &=& \frac{9 (1+\sum_i w_i\Omega_i)}{a^2 E^2 (1-3\sum_i w_i\Omega_i)^2} \left[\big( 1+\sum_i w_i\Omega_i\bigr)\bigl(1-3\sum_i w_i\Omega_i\bigr)\bigl(2+3\sum_i w_i\Omega_i\bigr) \right.\nonumber \\
    &{}& \left. +a \bigl(\frac{1}{3}\Omega_r^{\prime} +w_x\Omega_x^{\prime}+ w_x^{\prime}\Omega_x\bigr)\bigl(5-3\sum_i w_i\Omega_i\bigr)\right] \nonumber\\
    &=& \frac{S_{\text{tot}}^\prime}{S_{h0}}\left[\frac{\left(2+3\sum_i w_i\Omega_i\right)}{a} +\frac{ \left(5-3\sum_i w_i\Omega_i\right)(\frac{1}{3}\Omega_r^{\prime} +w_x\Omega_x^{\prime}+ w_x^{\prime}\Omega_x)}{(1+\sum_i w_i\Omega_i)(1-3\sum_i w_i\Omega_i)}\right].  
\end{eqnarray}

Both equations, \eqref{dS tot} and \eqref{ddS tot}, demonstrate that variations in the total entropy depend on the evolution and rate of change of the density parameters ($\Omega_i$ and $\Omega_i^{\prime}$) and the equations of state ($w_i$ and $w_i^{\prime}$) of all cosmic fluids contributing to the universe's total energy budget.  
Among these, the most significant contributions come from the horizon and dark energy components, with the latter playing a crucial role in satisfying the GSL in the late universe.  
It should also be noted that, as dark energy is the only fluid with a negative EoS, it is uniquely capable of driving $S^{\prime}_{\text{tot}}(a)$ to negative values.  
The interplay between the contributions from dark energy and the horizon entropy ultimately determines whether the total entropy increases or decreases over the course of cosmic evolution.\\

\textbf{Summary of cosmological thermodynamic equations.}
The set of equations describing the cosmological thermodynamic quantities includes the horizon entropy \eqref{S hor}, the Kodama–Hayward (horizon) temperature \eqref{T horizon}, the first \eqref{dS hor} and second \eqref{ddS hor} derivatives of the horizon entropy, the first \eqref{dS in} and second \eqref{ddS in} derivatives of the internal entropy, and finally, the first \eqref{dS tot} and second \eqref{ddS tot} derivatives of the total entropy.
This set of equations is summarized in Box \ref{box:thermo_eqs}.
For mathematical simplicity, we have divided all entropy-related expressions by $S_{h0}$ to obtain dimensionless quantities, although only $S_h$ is explicitly normalised.  
In our study, the numerical values of each quantity are not of primary importance; rather, it is their general behaviour throughout cosmic evolution that holds significance.  
In the specialised literature on this subject, various approaches are used to present cosmological thermodynamic quantities. 
In this study, we have chosen to express these quantities in a manner that explicitly reveals their dependence on cosmic evolution, specifically through the Hubble function $E$, the evolution of the material content $\sum_i w_i\Omega_i$, and the dynamics of the material content $(\sum_i w_i\Omega_i)' = \frac{1}{3}\Omega_r^{\prime} + w_x\Omega_x^{\prime} + w_x^{\prime}\Omega_x$.  
This formulation clearly highlights how each cosmological component individually contributes to the evolution of the total entropy, thus providing insight into the physical viability of the models from a thermodynamic perspective.

\renewcommand{\figurename}{Box}
\setcounter{figure}{0}  
\begin{figure}[h]
    \centering
    {\small
    \[
    \boxed{
    \begin{aligned}
        &\text{horizon temperature} && \frac{T_h}{T_{h0}} = E\left[1-\frac{3}{4}\bigl(1 +\sum_i w_i\Omega_i \bigr)\right] \\
        &\text{horizon entropy} && \frac{S_h}{S_{h0}} = \frac{1}{E^2} \\
        &\text{1st derivative of horizon entropy} && \frac{S^{\prime}_h}{S_{h0}} = \frac{3}{a E^2} \bigl(1 +\sum_{i}{w_i \Omega_i}\bigr) \\
        &\text{2nd derivative of horizon entropy} && \frac{S^{\prime \prime}_h}{S_{h0}} = \frac{S^{\prime}_h}{S_{h0}}\left[\frac{2 +3\sum_{i}{w_i \Omega_i}}{a} +\frac{\frac{1}{3}\Omega_r^{\prime} +w_x\Omega_x^{\prime}+ w_x^{\prime}\Omega_x}{1 +\sum_{i}{w_i \Omega_i}}\right] \\
        \hline
        &\text{1st derivative of internal entropy} && \frac{S_{\text{in}}^\prime}{S_{h0}} = -\left(\frac{S_h^\prime}{S_{h0}}\right) \left[1 - \frac{3\left(1 + \sum_i w_i \Omega_i \right)}{1 - 3\sum_i w_i \Omega_i} \right] \\
        &\text{2nd derivative of internal entropy} && \frac{S_{\text{in}}^{\prime\prime}(a)}{S_{h0}} = -\left(\frac{S_h^{\prime\prime}}{S_{h0}}\right) \left[1 - \frac{3\left(1 + \sum_i w_i \Omega_i \right)}{1 - 3\sum_i w_i \Omega_i} \right] 
        + 12 \left(\frac{S_h^\prime}{S_{h0}}\right) \left[\frac{\frac{1}{3}\Omega_r^\prime + w_x \Omega_x^\prime + w_x^\prime \Omega_x}{(1 - 3\sum_i w_i \Omega_i)^2}\right] \\
        \hline
        &\text{1st derivative of total entropy} && \frac{S_{\text{tot}}^\prime}{S_{h0}} = \frac{9}{a E^2}\frac{(1+\sum_i w_i\Omega_i)^2}{1-3\sum_i w_i\Omega_i} \\
        &\text{2nd derivative of total entropy} && \frac{S_{\text{tot}}^{\prime\prime}}{S_{h0}} = \frac{S_{\text{tot}}^\prime}{S_{h0}}\left[\frac{\left(2+3\sum_i w_i\Omega_i\right)}{a} +\frac{ \left(5-3\sum_i w_i\Omega_i\right)(\frac{1}{3}\Omega_r^{\prime} +w_x\Omega_x^{\prime}+ w_x^{\prime}\Omega_x)}{(1+\sum_i w_i\Omega_i)(1-3\sum_i w_i\Omega_i)}\right]
    \end{aligned}
    }
    \]
    }
    \caption{Set of key cosmological thermodynamic equations.}
    \label{box:thermo_eqs}
\end{figure}
\renewcommand{\figurename}{Figure}
\addtocounter{figure}{0} 

\section{$\Lambda$CDM thermodynamics}\label{sec:LCDM_thermo}

\begin{figure}[ht!]
    \centering
    \includegraphics[width=1.0\linewidth]{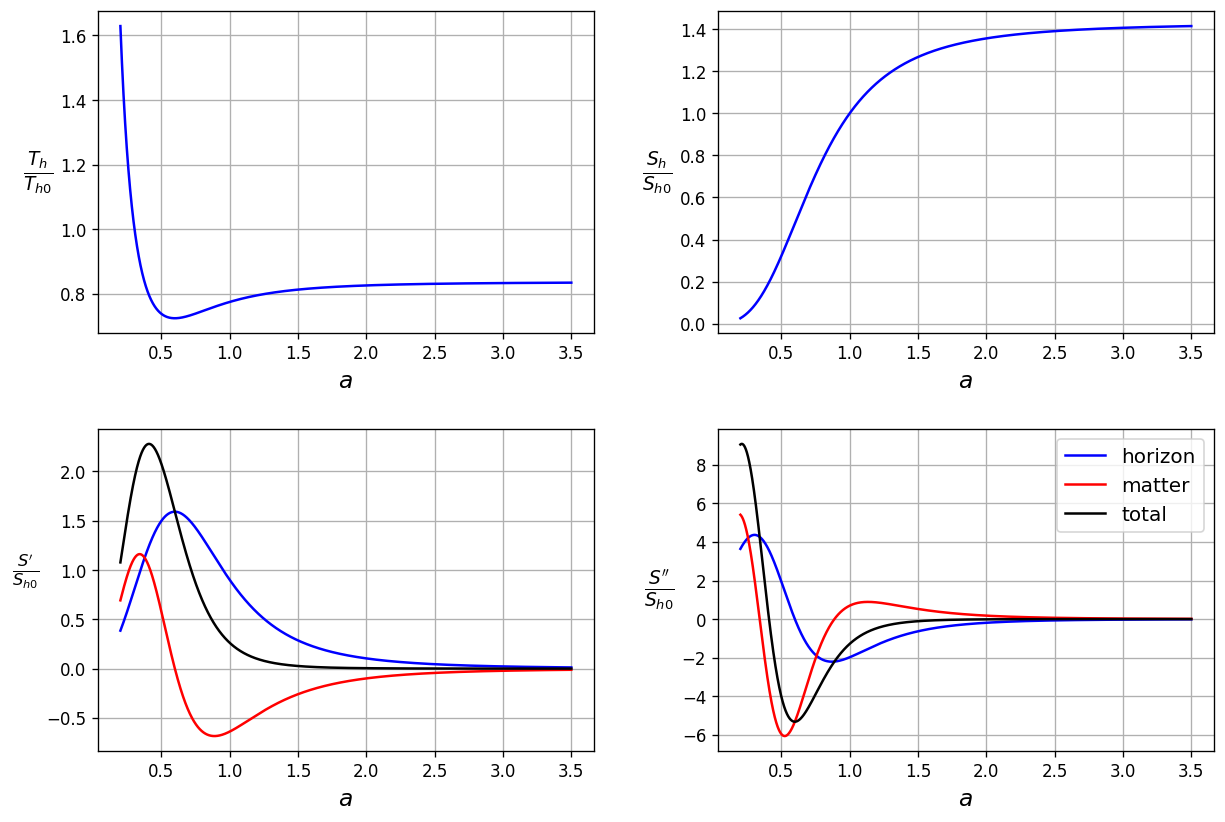}
    \caption{Evolution of the thermodynamic quantities in the $\Lambda$CDM model. The top left panel shows the horizon temperature, $T_h$ while in the top right panel the horizon entropy, $T_h$. The bottom left panel displays the first derivative of the total entropy, $S'_{\text{tot}}$ (black), with contributions from the horizon entropy, $S'_h$ (blue), and the internal entropy, $S'_{\text{in}}$ (red). The bottom right panel shows the second derivative of entropy, $S''_{\text{tot}}$ (black), along with $S''_h$ (blue) and $S''_{\text{in}}$ (red).}
    \label{fig:LCDM}
\end{figure}

As a starting point, the set of equations governing the cosmological thermodynamic quantities shown in Box \ref{box:thermo_eqs} is solved for the flat $\Lambda$CDM universe. 
We will first provide a detailed description of the thermodynamic quantities in $\Lambda$CDM. 
This serves two purposes: first, as the reference model for comparison with the sign-switching dark energy models, and second, because, as we will later see, the models under study generally exhibit similar behaviour to $\Lambda$CDM throughout cosmic history. 
The main exceptions occur during a brief period around the sign-switching moment at $a_*$ and in some asymptotic values in the far future for gDE. 
The results for the flat $\Lambda$CDM model are presented in Fig. \ref{fig:LCDM}.

\textbf{Horizon temperature.} (Fig. \ref{fig:LCDM} top left) 
The horizon temperature decreases rapidly, reaching a minimum around $a \approx 0.6$, before gradually rising to a stable asymptotic value.
From $a \approx 2$ onwards, it attains a constant value of $\frac{T_h}{T_{h0}}(a \gg 1) = \sqrt{\Omega_{x0}} \approx 0.86$. 
In the late-time limit, when the cosmological constant dominates cosmic expansion (the Hubble parameter approaches a constant), both horizon temperature and entropy stabilise at constant values.

\textbf{Horizon entropy.} (Fig. \ref{fig:LCDM} top right) 
Exhibits an initial phase of rapid growth, followed by a gradual reduction in its growth rate, eventually stabilising and asymptotically approaching a constant value for $a \gtrsim 2.5$. 
This behaviour is consistent with expectations.
The horizon entropy is inversely proportional to the square of the dimensionless Hubble parameter, $\frac{S_h}{S_{h0}} = \frac{1}{E^2}$, in the late universe where the cosmological constant dominates over the other components the Hubble parameter approaches a constant value, leading the horizon entropy to stabilise as well. 
Specifically, the entropy tends to $\frac{S_h}{S_{h0}}(a \gg 1) = \frac{1}{\Omega_{x0}} \approx 1.43$.
    
\textbf{Entropy first derivative.} (Fig. \ref{fig:LCDM} bottom left) 
$S'_h$ (blue), initially increases, reaching a maximum at $a \approx 0.6$, reflecting the significant expansion of the apparent horizon during the early universe. Then, it decreases but remains positive, asymptotically approaching zero for $a \gtrsim 3$, as the horizon’s growth stabilises during the dark energy-dominated era, where the expansion becomes nearly exponential, and the apparent horizon remains approximately constant.
$S'_{\text{in}}$ (red), rises rapidly, peaking at $a \approx 0.35$ during the matter-dominated phase, driven by efficient entropy production linked to structure formation and high matter density. 
It then decreases, reaching zero at $a \approx 0.6$, indicating a balance between entropy production and dilution due to cosmic expansion. 
Notably, beyond $a \approx 0.6$, $S'_{\text{in}}$ turns negative, signifying a period in which the internal entropy decreases.
After reaching a minimum at $a \approx 0.9$, $S'_{\text{in}}$ gradually increases, stabilising at negligible values from $a \gtrsim 3$.
In the far future, both $S'_{\text{in}}$ and $S'_h$ tend to zero, indicating a dynamic equilibrium where the total entropy is constant.
Finally, $S'_{\text{tot}}$ (black), consistently satisfies the GSL: $S'_{\text{tot}} \geq 0$ \eqref{S' tot}. 
It increases rapidly, peaking at $a \approx 0.4$, then decreases steadily, approaching zero asymptotically for $a \gtrsim 1.8$. 
Crucially, despite the negative values of $S'{\text{in}}$, the total entropy derivative $S'{\text{tot}}$ remains positive due to the dominant positive contribution from $S'_h$, thereby consistently satisfying the GSL.
    
\textbf{Entropy second derivative.} (Fig. \ref{fig:LCDM} bottom right) 
$S''_{\text{tot}}$ starts from a positive maximum at $a \approx 0.25$, during the matter-dominated era, reflecting efficient entropy production due to structure formation. 
It rapidly decreases, crossing zero around $a \approx 0.4$, marking the shift from an accelerating to a decelerating rate of entropy growth, closely associated with the onset of dark energy dominance.
Then reaches a negative minimum at $a \approx 0.6$, indicating maximum concavity corresponding to the increasing dominance of accelerated expansion, thus diminishing the efficiency of entropy production. 
Beyond this point, $S''_{\text{tot}}$ increases slowly, approaching zero around $a \approx 1.5$, as the universe enters a dark energy-dominated phase with the entropy stabilising in line with thermodynamic equilibrium. 
For $a \gg 1$, it asymptotically tends to zero, consistent with the final de Sitter-like stage, where entropy production effectively ceases.
$S''_h$, exhibits a similar qualitative behaviour to $S''_{\text{tot}}$, with its maximum and minimum occurring at larger scale factors, maximum at $a \approx 0.35$ and minimum at $a \approx 0.75$.  
In contrast, $S''_{\text{in}}$ exhibits a qualitatively similar pattern but with a notable difference. 
After reaching its negative minimum around $a \approx 0.5$, it increases, becomes positive, and reaches a maximum at $\frac{S''_{\text{in}}}{S_{h0}}(a \approx 1.1) \approx 1$, indicating a transient period of accelerated internal entropy growth, arising from the interplay between matter dilution and horizon evolution during the matter-to-dark-energy transition
After this peak, the function decreases and stabilises near zero around $a \approx 2.5$, confirming the system’s gradual approach towards thermodynamic equilibrium.
These results are consistent with the GSL condition $S''_{\text{tot}}(a) \leq 0$ for $a \gg 1$ \eqref{S'' tot}.
The changes in sign observed for $S''{\text{tot}}$ indicate inflection points in entropy growth dynamics and do not constitute violations of the GSL, as the first derivative, $S'{\text{tot}}$, remains non-negative at all times.
 
\section{Sign-switching dark energy thermodynamics}\label{sec:sign_switching_thermo}

In this section, we apply the previously described procedure to the three sign-switching dark energy models. 
While their thermodynamic evolution is generally similar to that of $\Lambda$CDM, as outlined in the previous section, notable deviations occur near the sign change at $a_*$ and, in some cases, in the distant future for gDE.  
Therefore, our analysis will focus on these differences, omitting aspects that align with $\Lambda$CDM. 

\subsection{Temperature, entropy, and entropy derivatives}

\begin{figure}[ht!]
    \centering
    \includegraphics[width=1.0\linewidth]{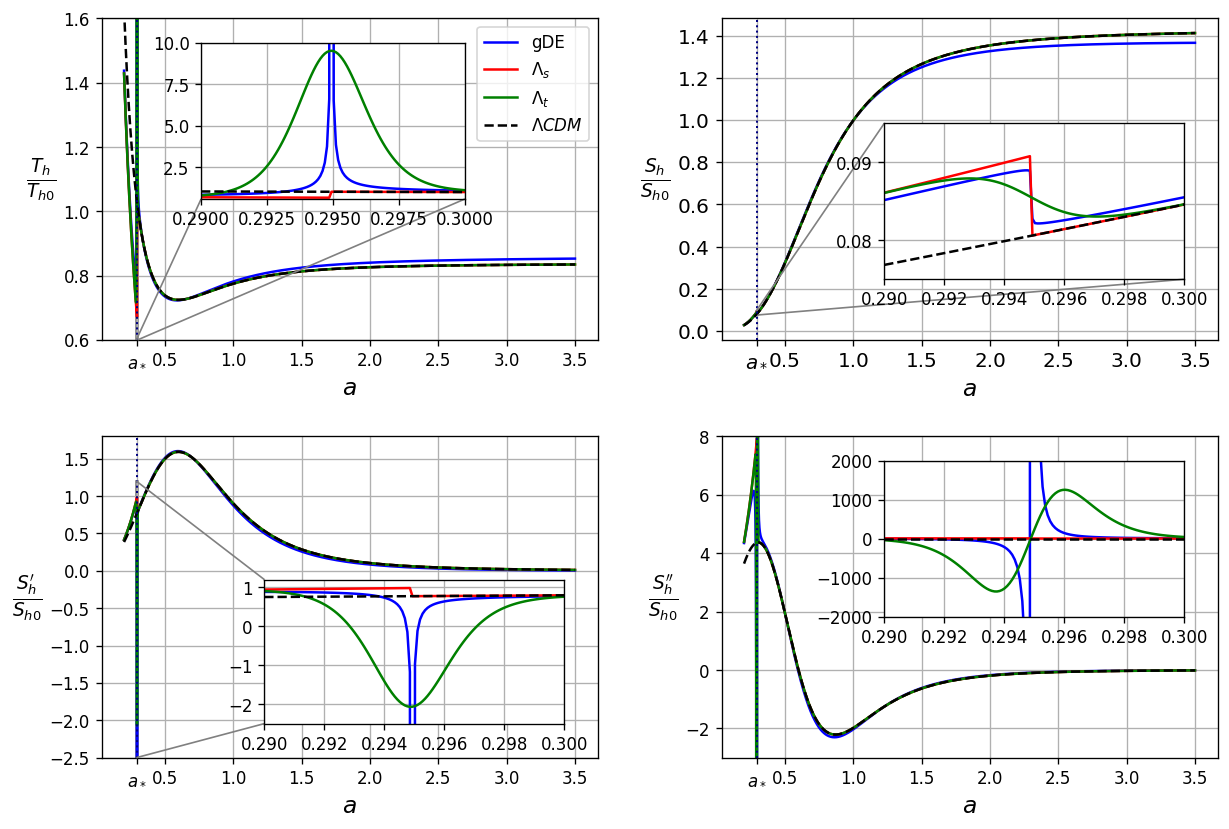}
    \caption{Evolution of the horizon temperature, horizon entropy and its first and second derivatives for gDE (blue), $\Lambda_s$ (red), and $\Lambda_t$ (green), with $\Lambda$CDM (dashed black) shown as a reference.}
    \label{fig:horizon}
\end{figure}

\textbf{Horizon temperature.} 
(Fig. \ref{fig:horizon} top left)  
The three sign-switching dark energy models initially decrease, starting at slightly lower values than $\Lambda$CDM.  
For gDE, very close to the transition, the temperature rises sharply, diverging at $a_*$ before dropping just as abruptly, then remains close to $\Lambda$CDM, but for $a \gtrsim 1$, it becomes higher than in the other models in the late future.  
The $\Lambda_t$ model also exhibits a sharp increase before the transition, though less pronounced than in gDE, it reaches a maximum at $a_*$ (approximately 10 times the corresponding $\Lambda$CDM value at that point) before decreasing and aligning with $\Lambda$CDM.  
In the sign-switching $\Lambda_s$ model, the transition is abrupt, after which it follows $\Lambda$CDM.  
The divergence of the horizon temperature in gDE is particularly striking and will be examined in a later section.

\textbf{Horizon entropy.} 
(Fig. \ref{fig:horizon} top right)   
$\Lambda_s$ and $\Lambda_t$ start with slightly higher values than $\Lambda$CDM but rapidly converge to it after the transition.  
The evolution of gDE differs from that of the other two models: initially, it exhibits slightly lower values, then, after the transition, it slightly exceeds $\Lambda$CDM before nearly aligning with it for a period; however, beyond $a \approx 1.1$, it decreases again, remaining lower than the other models and never approaching the $\Lambda$CDM asymptotic value of $\frac{S_h}{S_{h0}}(a \gg 1) = \frac{1}{\Omega_{x0}} \approx 1.43$, suggesting that the gDE model does not fully settle into a thermodynamic equilibrium at late times as efficiently as $\Lambda$CDM.

\textbf{Horizon entropy first derivative.}
(Fig. \ref{fig:horizon} bottom left)   
Overall, the sign-switching dark energy models exhibit an initial growth phase, reaching a maximum at $a \approx 0.6$, followed by a decline at a similar rate, and eventually becoming negligible for $a \gtrsim 3$.  
All three models start with higher values than $\Lambda$CDM, indicating greater entropy production compared to the standard model. 
Shortly before the transition, gDE and $\Lambda_t$ experience a sharp drop (more pronounced in gDE) entering the negative regime, signifying a temporary phase of entropy reduction due to rapid variations in dark energy density around the sign-switch transition.
At $a_*$, gDE diverges to negative infinity, while $\Lambda_t$ reaches a minimum before both recover and follow the standard evolution. 
In contrast, $\Lambda_s$ immediately aligns with $\Lambda$CDM after the transition. 
As with the horizon temperature, the most striking feature is the divergence observed in gDE.

\textbf{Horizon entropy second derivative.}
(Fig. \ref{fig:horizon} bottom right)  
The gDE model starts slightly below $\Lambda$CDM, then decreases, diverging to negative infinity before abruptly switching to positive infinity at $a_*$. 
It then rapidly drops and aligns with $\Lambda$CDM, all within a relatively short interval around $a_*$.  
The $\Lambda_t$ model also begins slightly lower than $\Lambda$CDM and undergoes a sharp oscillation around $a_*$, first reaching large negative values before surging to large positive values, ultimately settling to match $\Lambda$CDM.  
Unlike the other two models, $\Lambda_s$ starts higher than $\Lambda$CDM and aligns with it at the transition.  
In all cases, shortly after $a_*$, the models follow the behaviour of $\Lambda$CDM, tending to zero for $a \gg 1$.

\begin{figure}[ht!]
    \centering
    \includegraphics[width=1.0\linewidth]{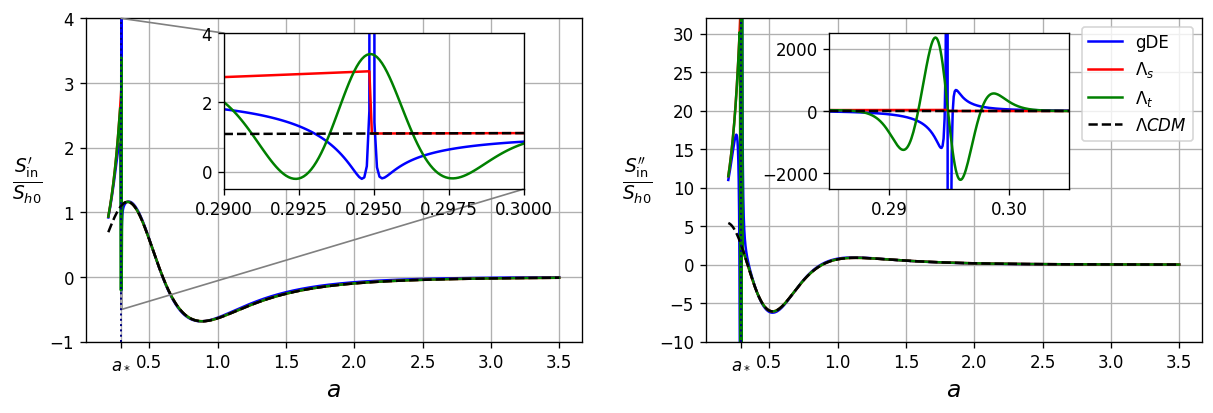}
    \includegraphics[width=1.0\linewidth]{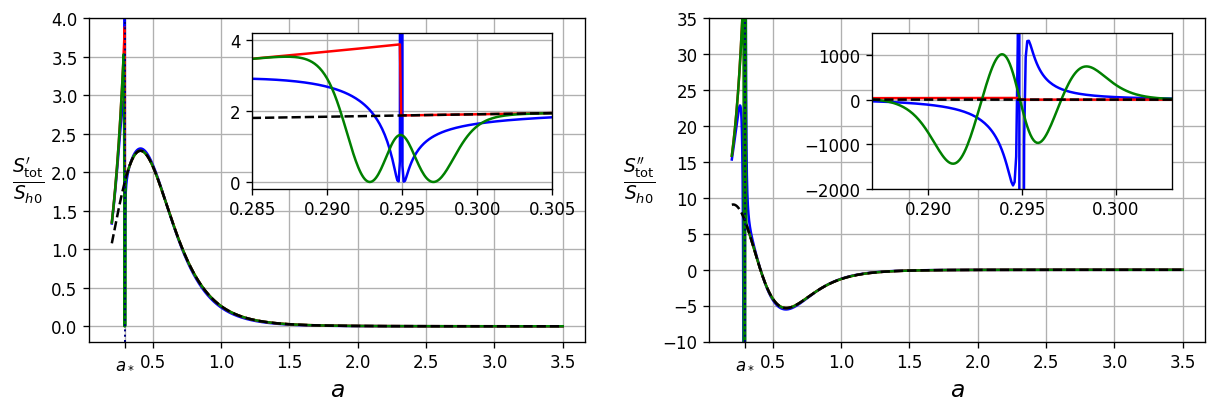}
    \caption{Evolution of the first and second derivatives of the internal entropy (top) and the total entropy (bottom) for gDE (blue), $\Lambda_s$ (red), and $\Lambda_t$ (green), with $\Lambda$CDM (dashed black) as a reference.}
    \label{fig:matter-total}
\end{figure}

\textbf{Internal entropy first derivative.}
(Fig. \ref{fig:matter-total} top right)
All three models start with values higher than $\Lambda$CDM.  
The gDE model decreases, even reaching negative values, before rising abruptly and diverging at $a_*$. 
It then undergoes a sharp decline and gradually approaches $\Lambda$CDM from below.  
The $\Lambda_t$ model also decreases and exhibits oscillations, becoming negative shortly before and after $a_*$ and reaching a maximum at the transition, then converges towards $\Lambda$CDM from below.  
The $\Lambda_s$ model aligns with $\Lambda$CDM at the transition.  
The most notable feature is that both gDE and $\Lambda_t$ exhibit behaviour opposite to that of $S'_h$, as they reach their maximum values at $a_*$, whereas $S'_h$ reaches its minimum.

\textbf{Internal entropy second derivative.}
(Fig. \ref{fig:matter-total} top left) 
Here, the behaviour of gDE and $\Lambda_t$ exhibits significant divergences and oscillations near the transition, indicating complex entropy dynamics.
Both start with values higher than $\Lambda$CDM and undergo oscillations with large amplitudes around $a_*$, where gDE exhibits both positive and negative divergences, while $\Lambda_t$ reaches distinct maxima and minima. 
Shortly after the transition, both converge towards $\Lambda$CDM.  
The $\Lambda_s$ model aligns with $\Lambda$CDM at the transition.

\textbf{Total entropy first derivative.}
(Fig. \ref{fig:matter-total} bottom left)
gDE starts with values higher than $\Lambda$CDM, as it approaches $a_*$, it decreases, reaching zero, then rises abruptly and diverges at $a_*$, it then undergoes a sharp decline, becoming zero again, before increasing and converging towards $\Lambda$CDM from below.  
The $\Lambda_t$ model follows a similar pattern: it starts above $\Lambda$CDM, decreases until it reaches zero, then rises to a local maximum at $a_*$, subsequently declines back to zero, and finally increases, converging towards $\Lambda$CDM from below.  
The $\Lambda_s$ model aligns with $\Lambda$CDM at the transition.  
Notably, all three models satisfy $S'_{\text{tot}} \geq 0$, even in cases where gDE and $\Lambda_t$ temporarily exhibit $S'_h < 0$ and $S'_{\text{in}} < 0$.  
This indicates that despite localised entropy reductions within individual components, the global thermodynamic evolution consistently adheres to the GSL.
The most striking feature that persists is the positive divergence of gDE.

\textbf{Total entropy second derivative.}
(Fig. \ref{fig:matter-total} bottom right)
The qualitative behaviour of $S''{\text{tot}}$ closely mirrors that of $S''{\text{in}}$, reinforcing the conclusion that internal entropy dynamics dominate the second derivative of the total entropy around the transition.

\subsection{Entropy derivatives of gDE, $\Lambda_s$ and $\Lambda_t$}

To gain deeper insights and refine the analysis, we now conduct a more detailed examination of the first and second derivatives of entropy (horizon, matter, and total) for each sign-switching dark energy model.  
This approach allows us to identify shared features and to clearly distinguish the specific thermodynamic behaviours of each model. 

\begin{itemize}
    \item In all cases, the most striking phenomena occur around $a_*$.  
    \item Near this point, the gDE model exhibits divergences, $\Lambda_s$ presents discontinuities, and $\Lambda_t$ undergoes high-amplitude oscillations.
    \item Highlight that the condition $S'{\text{tot}} \geq 0$ holds in all scenarios, despite the presence of discontinuities, divergences, or temporary negative values in $S'{\text{h}}$ and $S'_{\text{in}}$.
    \item For $a \approx 0.6$, $S'_{\text{in}}$ becomes negative and remains so, asymptotically approaching zero. This indicates a decrease in the entropy of matter. However, this decline is compensated by the growth of $S'_{\text{h}}$, ensuring that the condition $S'_{\text{tot}} \geq 0$ is maintained.
    \item The evolution of $S'_{\text{tot}}$ is primarily governed by $S'_{\text{in}}$, rather than $S'_{\text{h}}$, as evidenced by the fact that $S'_{\text{tot}}$ closely mirrors the behaviour of $S'_{\text{in}}$.
    \item Both $S'$ and $S''$ asymptotically approach zero: $S'{\text{h}}$ and $S''{\text{h}}$ decrease from above and below, respectively, while $S'{\text{in}}$ and $S''{\text{in}}$ increase from below and above, respectively.
    \item For $a \approx 0.4$, the condition $S''_{\text{tot}} \leq 0$ holds.
\end{itemize}

\begin{figure}[ht!]
    \centering
    \includegraphics[width=1.0\linewidth]{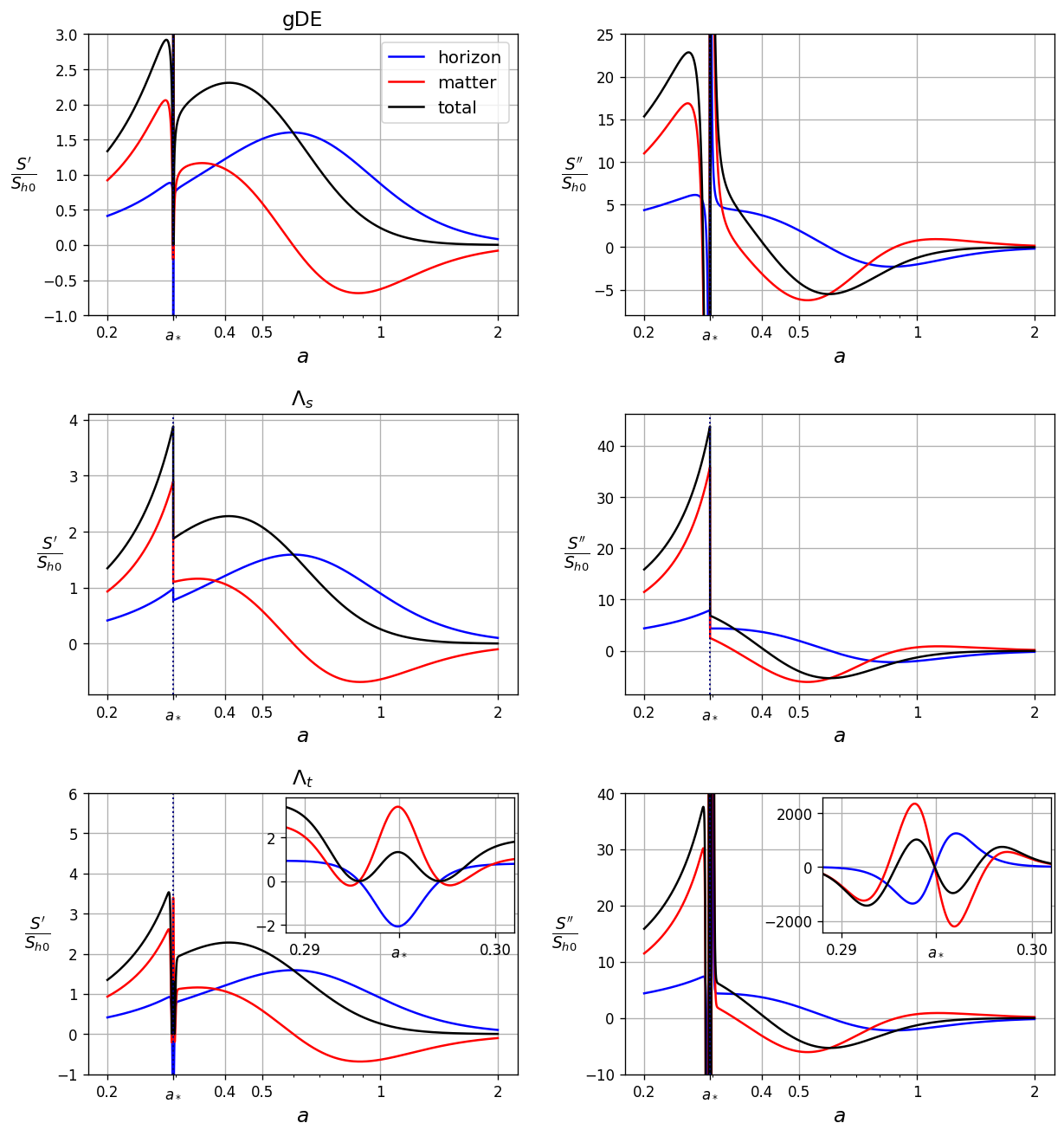}
    \caption{Evolution of the first and second derivatives of the horizon entropy (blue), internal entropy (red), and total entropy (black) for gDE (top), $\Lambda_s$ (middle), and $\Lambda_t$ (bottom).}
    \label{fig:dS-ddS}
\end{figure}

\textbf{gDE entropy derivatives.}
(Fig. \ref{fig:dS-ddS} top) 
At $a_*$, $S'_{\text{h}}$ diverges to negative infinity, while $S'_{\text{in}}$ diverges to positive infinity. 
When summed to form $S'{\text{tot}}$, the negative divergence from $S'{\text{h}}$ is outweighed by the positive divergence from $S'{\text{in}}$, ensuring that $S'{\text{tot}}$ remains positive. 
For the second derivatives, $S''_{\text{h}}$ initially diverges to negative infinity, then transitions to positive infinity before stabilising at finite values and asymptotically approaching zero.
Conversely, $S''_{\text{in}}$ and $S''_{\text{tot}}$ exhibit the opposite behaviour: they first diverge to positive infinity and subsequently to negative infinity.  

\textbf{$\Lambda_s$ entropy derivatives.}
(Fig. \ref{fig:dS-ddS} middle)
Here, both $S'$ and $S''$ increase rapidly, with $S''$ exhibiting a more pronounced growth before abruptly dropping at the discontinuous transition. Afterward, their behaviour aligns with that of the $\Lambda$CDM model.

\textbf{$\Lambda_t$ entropy derivatives.}
(Fig. \ref{fig:dS-ddS} bottom)
Oscillations occur around $a_*$.  
For $S'$, at $a_*$, $S'_{\text{in}}$ reaches the peak of its oscillation, while $S'_{\text{h}}$ attains a minimum.  
For $S''$, high-amplitude oscillations are observed. At $a_*$, both $S'_{\text{h}}$ and $S'_{\text{in}}$ converge to zero.  
Around the transition, $S'_{\text{h}}$ first reaches a local minimum, followed by a local maximum, whereas $S'_{\text{in}}$ exhibits the opposite behaviour.

\subsection{Divergences and asymptotic values of thermodynamic magnitudes}

\textbf{Divergences.} 
One of the most striking features of gDE is the emergence of divergences in thermodynamic quantities. 
To understand their origin, we first revisit the definition of horizon temperature, which was introduced using the Kodama-Hayward temperature \eqref{T horizon}, 
$\frac{T_h}{T_{h0}} = \frac{H}{H_0} (1 +\frac{\dot{H}}{2H^2}) = E[1-\frac{3}{4}(1 +\sum_i w_i\Omega_i)]$.
Since $H$ remains finite in the studied models (see the third panel of Fig. \ref{fig:rho w H}), any divergence must originate from the $\dot{H}$ term in the second Friedmann equation \eqref{Friedmann 2}.
This indicates that the divergence arises when cosmic acceleration ($\ddot{a}$) undergoes an abrupt and extreme variation.  
Previously, we rewrote the second Friedmann equation in a convenient form \eqref{friedmann 2 a}, which shows that the divergence stems from the term $\sum_i w_i\Omega_i$. 
This term is also present in all expressions for the entropy derivatives (horizon, internal, and total), as seen in Box \ref{box:thermo_eqs}.  
Since $w_r \Omega_r$ and $w_m \Omega_m$ remain regular, and that $ w_x \Omega_x = w_x \frac{\Omega_{x0} F_x}{E^2}$ we can immediately identify that the divergence originates from the product $w_x F_x$.  
From the second panel of Fig. \ref{fig:rho w H}, we see that the EoS $w_x$ diverges for both gDE and $\Lambda_t$. 
However, it is important to highlight that, in the case of $\Lambda_t$, this divergence in the EoS is cancelled in the product $w_x F_x$, thus preventing divergences in the thermodynamic quantities of the $\Lambda_t$ model.  

The fact that the horizon temperature and entropy production become infinite at the transition suggests two possible interpretations: 
First, it may indicate that the thermodynamic description of the horizon momentarily breaks down, implying that the usual assumptions underlying the derivation of $T_h$ and $S'$ may not hold at $a_*$. 
Second, as by construction, the gDE model induces a divergence in the EoS and given the functional form of the horizon temperature and the first and second derivatives of the horizon entropy, which directly depend on $w_x$, these quantities inevitably inherit the divergence.  
This suggests a fundamental thermodynamic inconsistency in the gDE model, which could undermine its physical plausibility despite its observational successes in alleviating cosmological tensions.

\textbf{Asymptotic values.}
On the other hand, the asymptotic values for the distant future of the physical quantities also provide valuable insights. Let us revisit the forms of the sign-switching models, eqs. \eqref{rhox}, \eqref{rhox ss}, \eqref{rhox tahn}:
\begin{eqnarray}
    F_x(a) =
\begin{cases}
    \text{sgn}\left(1-\Psi \ln a\right) \left| 1-\Psi \ln a \right|^{\frac{1}{1-\lambda}}, & \text{for gDE} \\
    \text{sgn}\left(\frac{1}{a_*} - \frac{1}{a} \right), & \text{for } \Lambda_s \\
    \tanh\left[\eta\left(\frac{1}{a_*} - \frac{1}{a} \right)\right], & \text{for } \Lambda_t
\end{cases}
\end{eqnarray}
Note that, without introducing additional elements into the models, $\Lambda_s$ and $\Lambda_t$ naturally converge to $\Lambda$CDM in the future, meaning that $F_x(a \gg 1) = 1$.  
In contrast, for gDE, the term $|1-\Psi \ln a|^{\frac{1}{1-\lambda}}$ remains, as the exponent is small but positive, satisfying $1 > \frac{1}{1-\lambda} > 0$. Although its growth is extremely slow—given that it follows a logarithm raised to a small power—it is asymptotically divergent, which carries significant implications.
\begin{eqnarray}
    \lim_{a\rightarrow \infty} F_x(a) =
\begin{cases}
    \infty, & \text{for gDE} \\
    1. & \text{for } \Lambda_s, \Lambda_t 
\end{cases}
\end{eqnarray}
This means that the energy density of gDE continues to grow indefinitely, which implies:
\begin{eqnarray}
    \lim_{a\rightarrow \infty} E =
\begin{cases}
    \infty, & \text{for gDE} \\
    \sqrt{\Omega_{x0}}. & \text{for } \Lambda_s, \Lambda_t 
\end{cases}
\end{eqnarray}
Thus, we can determine the asymptotic values of the horizon temperature \eqref{T horizon} and the horizon entropy \eqref{S hor}.
\begin{align}
    \lim_{a\rightarrow \infty} \frac{T_h}{T_{h0}} &=
    \begin{cases}
        \infty, & \text{for gDE} \\
        \sqrt{\Omega_{x0}}, & \text{for } \Lambda_s, \Lambda_t
    \end{cases} \label{T hor asymp} \\
    \lim_{a\rightarrow \infty} \frac{S_h}{S_{h0}} &=
    \begin{cases}
        0, & \text{for gDE} \\
        \frac{1}{\Omega_{x0}}, & \text{for } \Lambda_s, \Lambda_t
    \end{cases} \label{s hor asymp}
\end{align}

In the distant future, $\Lambda_s$ and $\Lambda_t$ exhibit the same finite and constant horizon temperature and entropy values as $\Lambda$CDM, which immediately implies that their derivatives tend to zero. 

In contrast, for gDE, the horizon temperature asymptotically diverges while the entropy approaches zero. 
This leads to a peculiar scenario in which the temperature grows without bound, yet the entropy vanishes.  
In most physical systems, an increase in temperature is typically accompanied by an increase in entropy. 
This is well illustrated by ideal gases, where the entropy depends on temperature following the Sackur–Tetrode equation, $S \sim k_B \ln T$.  
Even more puzzling is the fact that, since the horizon entropy tends to zero from an initially positive value (see the top right panel of Fig. \ref{fig:horizon}), it must be decreasing in the far future. 
In other words, $S'_h < 0$ for $a \gg 1$, which constitutes a violation of the GSL. 

\subsection{Thermodynamic evaluation}

Under the assumption that the Universe can be treated as an isolated thermodynamic system bounded by the apparent horizon, the GSL imposes fundamental conditions on the entropy and its derivatives, as expressed in Section~\ref{sec: Cosmological thermodynamics}.
These criteria establish a robust theoretical framework against which any dark energy model can be evaluated, independently of its phenomenological motivation or observational success.  
Based on these principles, our thermodynamic evaluation yields the following key conclusions:  

\begin{itemize}
    \item \textbf{$\Lambda$CDM:}  
    The standard cosmological model fully satisfies all thermodynamic criteria imposed by the GSL. 
    The horizon entropy exhibits a smooth, monotonic growth throughout cosmic evolution. 
    The first derivative of the total entropy remains strictly positive, while its second derivative becomes negative and approaches zero at late times. 
    No divergences or discontinuities are observed in any thermodynamic quantity. 
    Both the horizon temperature and entropy converge to well-defined, finite asymptotic values, confirming the model’s full consistency with fundamental thermodynamic principles.
    
    \item \textbf{$\Lambda_s$ and $\Lambda_t$:}  
    Despite their non-standard construction, involving abrupt or smoothed sign-switching transitions, both the $\Lambda_s$ and $\Lambda_t$ models remain thermodynamically viable.
    Their deviations from the $\Lambda$CDM baseline are confined to the vicinity of the transition epoch, where $\Lambda_s$ displays discontinuities and $\Lambda_t$ undergoes pronounced oscillations.
    Importantly, neither model shows pathological behaviour such as divergences in thermodynamic quantities, even though $\Lambda_t$ features a divergent EoS parameter.
    The first and second derivatives of the total entropy consistently satisfy the GSL conditions, and their asymptotic evolution converges to that of the standard $\Lambda$CDM model.

    \item \textbf{gDE:}  
    In stark contrast, the gDE model exhibits significant thermodynamic inconsistencies.  
    The divergences in its EoS parameter, $w_x$, propagate into the horizon temperature and into the first and second derivatives of both horizon and internal entropies at the transition scale factor $a_*$.  
    These divergences render the thermodynamic quantities ill-defined in that region, indicating either a breakdown of the thermodynamic framework employed here or a fundamental inconsistency in the gDE model from a thermodynamic standpoint.  
    Moreover, in the asymptotic future, the horizon temperature diverges while the horizon entropy vanishes, resulting in a regime where $S'_h < 0$ for $a \gg 1$, in direct violation of the GSL.  
    Altogether, these issues call into question the thermodynamic consistency — and thus the physical viability — of the gDE model, despite its observational appeal.

    \item \textbf{General theoretical insight:}  
    Beyond the models studied in this work, we have demonstrated a significant theoretical insight: any dark energy model for which the product $w(a)\rho(a)$ diverges—typically due to divergences in the EoS parameter—will inevitably lead to divergences in the horizon temperature $T_h$, and the first and second entropy derivatives, $S'$ and $S''$, within the presented thermodynamic framework. 
    Such divergences strongly suggest incompatibility with the standard cosmological thermodynamic formulation, thus providing an important criterion to evaluate the physical viability of dark energy models beyond observational constraints alone.
\end{itemize}  

In summary, thermodynamic analysis provides an independent and fundamental criterion for assessing the viability of dark energy models. 
Models that may fit current observational data but violate fundamental thermodynamic laws, particularly the GSL, must be reconsidered. 
The framework presented here serves as a diagnostic tool, complementing observational constraints.

\section{Conclusions}\label{sec:conclusions} 

In this work, we have carried out a comprehensive thermodynamic analysis of sign-switching dark energy models, specifically focusing on the graduated dark energy (gDE), the abrupt sign-switching cosmological constant ($\Lambda_s$), and its smooth-transition variant described by a hyperbolic tangent function ($\Lambda_t$), within the context of a spatially flat FLRW universe. 
Our main objective was to evaluate the thermodynamic viability of these models beyond their observational effectiveness in addressing cosmological tensions. 
To achieve this, we began by providing a clear, systematic, and fully general formulation of the essential cosmological thermodynamic quantities, including the horizon temperature, horizon entropy, internal entropy, total entropy, and their first and second derivatives. 
These expressions, summarized compactly in Box \ref{box:thermo_eqs}, are model-independent, relying solely on the flat FLRW framework. 
Their concise and transparent presentation explicitly highlights their dependence on the Hubble expansion rate and the universe’s energy content, facilitating physical interpretation. 
This formulation represents a significant methodological advancement that can be readily extended and applied to a variety of other cosmological scenarios.

We first applied this formalism to the standard $\Lambda$CDM model, reproducing well-known behaviours and verifying that the evolution of total entropy satisfies all the conditions imposed by the Generalised Second Law (GSL) of thermodynamics. 
This provided a baseline for comparison. 
Subsequently, we extended our analysis to three representative sign-switching dark energy models.
Our results show that both $\Lambda_s$ and $\Lambda_t$, despite their unconventional construction involving sign changes in the dark energy density, remain thermodynamically consistent. 
After the sign-switching transition, these models evolve similarly to $\Lambda$CDM, with horizon temperature, entropy, and their derivatives that tend toward finite, well-behaved asymptotic values.
Importantly, both models satisfy the GSL conditions.
In contrast, the gDE model exhibits significant thermodynamic inconsistencies. 
The divergence in its EoS parameter $w_x$ propagates directly into the thermodynamic quantities, causing the horizon temperature and the first and second derivatives of entropy to diverge at the transition. 
Even more problematic is its asymptotic future behaviour: the horizon temperature diverges while the horizon entropy asymptotically approaches zero, leading to a regime of decreasing entropy in the very distant future. 
This explicitly violates the GSL, raising serious questions regarding the fundamental physical viability of the gDE model, despite its success in fitting cosmological data and alleviating tensions.
Furthermore, beyond the specific models here, our analysis also reveals an important theoretical insight: divergences in the product of the dark energy EoS parameter and its density ($w_x \Omega_x$) strongly correlate with divergences in $T_h$, $S'$, and $S''$. 
This observation suggests that any dark energy model exhibiting such divergences is likely to face thermodynamic inconsistencies
Thus, our results provide a powerful diagnostic criterion for evaluating the physical plausibility of dark energy models, complementing observational constraints with essential thermodynamic considerations.

The significance of this work lies in demonstrating that thermodynamic consistency represents a fundamental and complementary criterion that any physically viable cosmological model must satisfy.
Observational agreement, though essential, is insufficient if a model violates basic thermodynamic principles.
The systematic approach developed here thus provides a robust diagnostic tool, capable of identifying and discarding models with intrinsic theoretical inconsistencies.
Future research directions within the context of sign-switching dark energy models include a detailed exploration of the potential link between dark energy density transitions and thermodynamic phase transitions. 
It would also be valuable to extend the current analysis to broader classes of dark energy models. 
Additionally, exploring the implications of non-extensive entropies within this thermodynamic framework represents a promising avenue for future research.
\textcolor{blue}{Although the models studied here do not exhibit a dynamical dark energy behavior at low redshift, this feature is intrinsic to their formulation and consistent with the aim of testing their thermodynamical rather than observational viability.
In this context, it would also be interesting to apply the thermodynamical framework developed here to more dynamical dark energy scenarios—such as the wXCDM \cite{Gomez-Valent:2024tdb} model—in order to examine whether their sign-switching and quintessence-like behaviors remain thermodynamically consistent, particularly in their asymptotic future evolution.}

\bibliography{References.bib}

\end{document}